\documentclass{article}
\usepackage{graphicx}
\usepackage{amsmath}
\usepackage{multirow}
\usepackage{authblk}

\title{Bimanual proprioception: are two hands better than one?}

\author[1,2]{Jeremy D Wong}
\author[2]{Elizabeth T Wilson}
\author[1,2]{Dinant A Kistemaker}
\author[1]{Paul L Gribble\thanks{paul [at] gribblelab [dot] org}}
\affil[1]{Brain and Mind Institute, Western University, London Ontario Canada N6A 5B7}
\affil[2]{MOVE Research Institute VU University Amsterdam, The Netherlands 1081BT}

\date{July, 2013}

\begin{document}

\maketitle

\begin{abstract}
  Information about the position of an object that is held in both
  hands, such as a golf club or a tennis racquet, is transmitted to
  the human central nervous system from peripheral sensors in both
  left and right arms. How does the brain combine these two sources of
  information? Using a robot to move participant's passive limbs, we
  performed psychophysical estimates of proprioceptive function for
  each limb independently, and again when subjects grasped the robot
  handle with both arms. We compared empirical estimates of bimanual
  proprioception to several models from the sensory integration
  literature: some that propose a combination of signals from the left
  and right arms (such as a Bayesian maximum-likelihood estimate), and
  some that propose using unimanual signals alone. Our results are
  consistent with the hypothesis that the nervous system both has
  knowledge of, and uses the limb with the best proprioceptive acuity
  for bimanual proprioception. Surprisingly, a Bayesian model that
  postulates optimal combination of sensory signals could not predict
  empirically observed bimanual acuity. These findings suggest that
  while the central nervous system seems to have information about the
  relative sensory acuity of each limb, it uses this information in a
  rather rudimentary fashion, essentially ignoring information from
  the less reliable limb.
\end{abstract}

\section*{Introduction}
The human sensorimotor system can combine multiple sensory signals to
estimate the position of the body. Several studies have shown data
consistent with the hypothesis that the sensory system optimally
integrates sensory information: both when combining a prior
distribution with current signal variability \cite{Kording:04a}, and
when integrating visual and haptic sensory information
\cite{Ernst:02}. These studies suggest that the central nervous system
may implement some form of Bayesian statistics.

Studies of information processing by the nervous system have found
Bayesian models to often be consistent with empirical data, for a
rather broad set of behaviours that includes infant cognition
\cite{Gweon:10}, language \cite{Bannard:09,Frank:12}, face perception
\cite{Peterson:12}, rhythm perception \cite{Cicchini:12}, haptics
\cite{Squeri:12}, and multi-signal integration \cite{Ernst:02}. It is
thus an important current theory for sensory-motor neuroscience and
motor control. In some of these studies however it is not clear that
the Bayesian proposal is unique. It has sometimes been the case that
Bayesian models have been applied without the capacity to distinguish
between subtle differences in the underlying variability distributions
\cite{Zhang:13}. Thus some behaviours might be mistakenly classified
as Bayesian as a result of mis-approximation of true sensory or motor
variability.

Proprioception of the human upper limb has been explored considerably
and some asymmetries between that of the dominant and non-dominant
limb have been observed \cite{Goble:08b}. In particular there is some
suggestion that the non-dominant arm may have superior position
sense. This is an ecologically relevant question because many
behaviours involve the simultaneous use of both hands and thus the
central nervous system may implement some form of sensory
integration. Here we directly test the nervous system's integration of
proprioceptive signals from the left and right arms.

In this experiment we perform psychophysical estimates of
proprioceptive function. By measuring unimanual proprioception of each
limb and comparing these measures with bimanual proprioception we can
test if and how the human sensorimotor system combines proprioceptive
signals from the two limbs. We compare empirically observed bimanual
proprioceptive bias and acuity to predictions from several models of
sensory integration, including some that propose combining signals
from the left and right arms, and some that propose using unimanual
signals alone. Our results are consistent with the hypothesis that the
nervous system both has knowledge of and uses the limb with the best
proprioceptive acuity for bimanual proprioception. Our data are not
consistent with the hypothesis that the sensorimotor system optimally
combines unimanual proprioceptive signals from the two limbs in the
bimanual case.

\section*{Materials and Methods}

\subsection*{Subjects}
37 (20 female) healthy individuals participated in this study (aged 18
to 45 years). Subjects reported no history of neurological or
musculoskeletal disorder, and had normal or corrected-to-normal
vision. All subjects provided written informed consent prior to
participation in the study, which was approved by the University of
Western Ontario Research Ethics Board.

\subsection*{Apparatus}
\begin{figure}[p]
  \begin{center}
    \includegraphics[height=14cm]{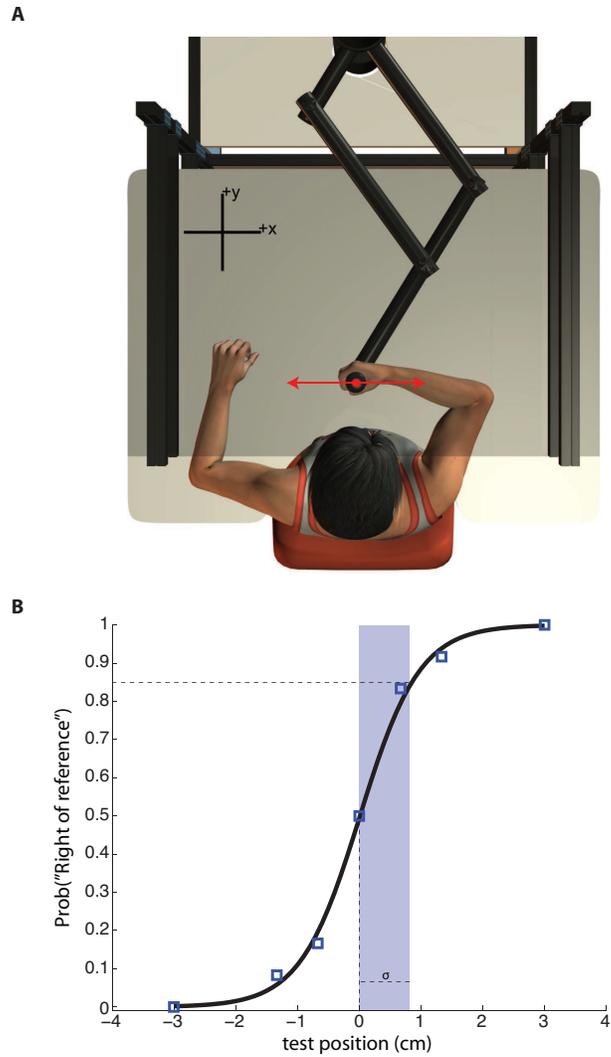}
   \end{center}
   \caption[Figure 1]{A: Subjects sat at a table and grasped the
     robotic manipulandum during proprioceptive tests of Left, Right
     and Bimanual judgments. B: Example Psychometric Function. Squares
     denote the probability with which a subject reported a given hand
     position to be right of the reference location, as a function of
     the actual hand location. Subjectsʼ responses were fit to using a
     cumulative normal distribution function. The vertical dashed line
     indicates the bias; here the estimated bias was 0. The shaded
     region represents the estimated proprioceptive acuity, $\sigma$,
     of 8.4~mm.}
 \label{fig:setup}
\end{figure}

Subjects were seated in the dark at a table adjusted to chest
height. Subjects grasped the handle of an InMotion robotic linkage (In
Motion Technologies, Cambridge, USA) as shown in
Figure~\ref{fig:setup}A. An air sled was used to support the arm and
allowed smooth, near frictionless movement along the surface of the
table (not shown). The robot was programmed to move the arm from one
position to another in a two-dimensional horizontal plane located just
below shoulder height. A six-axis force transducer (ATI Industrial
Automation, Apex, USA) inside the handle measured forces at the
hand. Shoulder straps attached to the chair kept the trunk in a static
position, while allowing rotation of the shoulder and elbow joints. A
horizontal semi-silvered mirror was suspended 31.5 cm above the
surface of the table. Vision of the arm and the robotic manipulandum
was obscured by opaque curtains in addition to the semi-silvered
mirror.

\subsection*{Proprioceptive tests}
Proprioceptive tests were performed at a single spatial location along
the sagittal plane 18 cm away from the body. Three tests were
performed in series by each subject: one in which they grasped the
handle of the robot using their left hand only, a second in which they
grasped the handle using the right hand only, and a third in which
they grasped the handle using both hands simultaneously. When testing
bimanual proprioception, the subject's fingers were interleaved such
that neither hand gripped the handle more directly than the other. The
order in which the tests were performed was counterbalanced across
subjects.

\subsection*{Psychophysical estimates of limb proprioception}
The test procedure has been described elsewhere
\cite{Wilson:10,Ostry:10,Wong:11}. Briefly, we employed a
two-alternative forced-choice paradigm to estimate the psychophysical
relationship between actual and perceived position of the limb(s). On
each trial, subjects were instructed to keep their arm muscles
relaxed, and their head in a neutral, forwards direction. Vision of
the arm was completely blocked by opaque curtains. Each proprioceptive
test consisted of 74 trials in which the robot moved the passive
limb(s) along a left-right axis.

Subjects were instructed to keep their eyes closed at all times. On
each trial, the subject's arm was moved to the reference position by
the robotic manipulandum, and held there for 2~s. Next, the hand was
moved away from the reference position through a distractor movement,
before being brought to a judgment position where the hand was held
until the subject made a two-alternative forced-choice judgement about
which side along the axis of movement (left or right) the judgment
position fell with respect to the reference position. The distractor
movement displaced the hand 14 cm plus or minus a random distance
(chosen from a gaussian with mean = 14 cm and sd = 2 cm) from the
reference position along the test axis to a peripheral position before
bringing the hand to a judgment position. Seven judgment positions
were tested: $[-30, - 13.3, -6.7, 0, +6.7, +13.3, +30]$~mm. Each
judgment position was tested between 6 and 14 times $[6, 12, 12, 14,
12, 12, 6]$. The positions furthest from the reference position were
tested fewer times because subjects were expected to make essentially
no errors at these distant positions.

To familiarize the subject with the procedure, blocks of 20 practice
trials were performed at the start of the experiment, until subjects
demonstrated a clear understanding of the task. The majority of
subjects only required a single practice block.

A logistic function was fit to the set of binary response data across
test locations (Fig.~\ref{fig:setup}B). Proprioceptive Bias was
quantified as the 50th percentile, i.e. the point at which subjects
were equally likely to report their hand as left or right of the
reference position. Proprioceptive acuity was quantified as $\sigma$,
the distance spanning the 50th to the 84th percentiles of the logistic
function. Statistical analysis of changes in proprioception were
assessed using analysis of variance and Tukey post hoc tests.

\subsection*{Predictions}
We tested several models, some proposed previously, that predict how
the central nervous system might use the two unimanual signals to
perform a bimanual estimate of hand position. These hypotheses can be
divided into those that predict signal combination and those that
propose the use of a single signal for perceptual judgments.

\subsubsection*{Signal combination models}

In this study the variance of a proprioceptive signal ($\sigma^{2}$)
was estimated using the square of the distance between the 50th and
84th percentile of the psychometric function. The reliability of a
given signal ($r$) is the inverse of the variance ($\sigma^{2}$):
\begin{equation}
  r = \frac{1}{\sigma^{2}}
\end{equation}

\paragraph*{Equal-weight}
The parsimonious signal-combination hypothesis predicts that the
reliability of the bimanual estimate $\hat{r}_{LR}$ is the average of
the two unimanual estimated reliabilities ${r}_{L}$ and ${r}_{R}$:
\begin{equation}
  \hat{r}_{LR} = \frac{r_R r_L}{r_R w^2 + r_L (1-w)^2} 
\end{equation}
where $w=\frac{1}{2}$. We refer to this prediction as $H_{BiEqual}$.

\paragraph*{Maximum likelihood estimation}
A second hypothesis predicts that the CNS optimally combines the
unimanual signals to generate a maximum likelihood estimate (MLE) of
bimanual position \cite{Ernst:02}. The MLE model predicts that the
bimanual estimate $\hat{r}_{LR}$ is a weighted combination of the two
unimanual estimates ${r}_{L}$ and ${r}_{R}$ for the left and right
arms. The weights are defined as a function of the unimanual
reliabilities:
\begin{align}
  w_{R} &= \frac{r_{R}}{r_{L} + r_{R}} \\
  w_{L} &= \frac{r_{L}}{r_{L} + r_{R}}
\end{align}
Thus less reliable unimanual estimates contribute to a bimanual
estimate with lower weight. The resulting variance of the bimanual
estimate ($\sigma^{2}$) is:
\begin{equation}
  \hat{\sigma}^{2}_{LR} = \frac{(\sigma^{2}_{L}) (\sigma^{2}_{R})}{\sigma^{2}_{L} + \sigma^{2}_{R}}
\end{equation}
For each subject in the experiment, we computed the predicted bias
($\hat{S}_{LR}$) and predicted acuity ($\sigma_{LR}$, the distance
between the 50th and 84th percentile of the psychophysical function)
under the MLE model and compared the predictions to the observed
values when subjects grasped the robot handle with both left and right
hands. This prediction was referred to as $H_{BiMLE}$.

\paragraph*{Maximum likelihood for correlated variables}
Finally, note that maximum likelihood estimation as described above
assumes uncorrelated signals. It might rather be the case that
unimanual signals are correlated. This might for example arise from
noise due to torso movement or within shared pathways in the central
nervous system.

To take into account the possibility of correlated signals, we did the
following. We assume a constant unknown degree of correlation $\rho$
between Left and Right unimanual signals across all subjects. We then
determined the correlation coefficient $\rho$ that resulted in best
fits of the observed Bimanual acuity according to the following
equation \cite{Oruc:03}:
\begin{equation}
  \hat{r}_{LR} = \frac{r_{L} + r_{R} - 2\rho\sqrt{r_{L}r_{R}}}{1-\rho^{2}}
\end{equation}
Using this estimate of $\rho$, we computed new a maximum likelihood
estimate. Including a correlation of $\rho$ between signals discounts
the predicted optimal Bimanual prediction, while having no effect on
the combined bias. This prediction was referred to as $H_{BiMLEcorr}$.

\subsubsection*{Single signal models}
Alternatively, the CNS might select a single limb for all perceptual
responses. Instead, the set of Bimanual responses might be generated
by
\begin{itemize}
\item use of the limb with the best proprioceptive acuity;
  $H_{UniMin}$
\item use of a single unimanual cue chosen at random for each trial;
  $H_{SwitchRand}$; or
\item use of a single unimanual cue chosen with probability
  proportional to the signal reliability; $H_{SwitchWeight}$
\end{itemize}	
To generate these last two predictions, we performed simulations of
the psychophysical experiments. Using the empirically estimated Left
and Right unimanual psychophysical curves of each subject, we
generated random-draw or weighted-draw responses that were used to
simulate bimanual responses. To generate predictions for
$H_{SwitchRand}$, we generated simulated responses where on each
trial, a binomial response was generated using either the Left or
Right empirically observed psychometric curve, chosen at random (with
equal probability) for each trial. To generate predictions for
$H_{SwitchWeight}$, the same procedure was used, except that instead
of basing the simulated responses on the Left or Right psychometric
functions chosen at random with equal probability, the probability of
using Left vs Right was proportional to signal reliability (the
inverse of acuity). After generating simulated responses for each
subject under both hypotheses, we re-estimated psychometric functions
and re-computed estimates of bias and acuity.
\section*{Results}

\subsection*{Proprioceptive bias}
Figure~\ref{fig:mf2}A (``empirical data'') shows mean $\pm$ standard
error (se) of the psychophysical estimates of perceptual bias for the
Right (blue), Left (red) and Bimanual (black) data, averaged across
subjects. Average $\pm$ standard error of $Bias_{Right}$ was -1.18
$\pm$ 0.41 ; $Bias_{Left}$ was 3.19 $\pm$ .36~mm; and
$Bias_{Bimanual}$ was 0.60 $\pm$ 0.34~mm. A split-plot
repeated-measures analysis of variance (one within-subjects variable,
Grasp [R,L,B] and one between-subjects variable, Testing Order [6
different permutations]) showed no main effect of order (p $>$ 0.05),
a significant main effect of grasp (p$<$0.001) and no interaction (p
$>$ 0.05); post-hoc tests showed significant differences between the
bias of all three conditions (p$<$0.001 in all pairwise
comparisons). Thus Left, Right and Bimanual biases were reliably
different from each other. Interestingly, $Bias_{Bimanual}$ was
between $Bias_{Right}$ and $Bias_{Left}$.

\subsection*{Proprioceptive acuity}
Figure~\ref{fig:mf2}B (``empirical data'') shows mean $\pm$ se of the
acuity, $\sigma$. Mean $\pm$ se acuity measures for Right, Left and
Bimanual are: 10.92 $\pm$ 0.55~mm, 12.23 $\pm$ 0.51~mm and 10.15 $\pm$
0.52~mm. A split-plot analysis of variance found a marginal main
effect of Grasp (p $=$ 0.066), no significant main effect of Order and
no interaction (p$>$ 0.05); post-hoc tests showed that
$Acuity_{Bimanual}$ was reliably different from $Acuity_{Left}$.

\subsection*{Model predictions for bias and acuity}
\begin{figure}[p]
  \begin{center}
    \includegraphics[height=15cm]{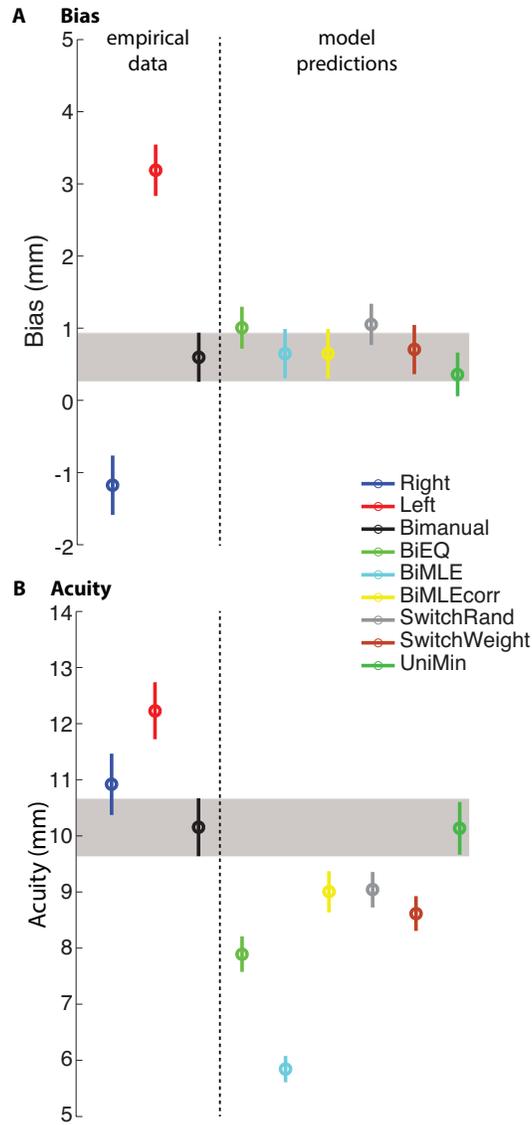}
   \end{center}
   \caption{A: Bias measures from empirical data (Right, Left and
     Bimanual) and predictions (Bi50,BiMLE, BiMLEcorr, SwitchRand,
     SwitchWeight, and UniMin). B: Acuity measures from empirical data
     (Right, Left and Bimanual) and predictions (Bi50,BiMLE,
     BiMLEcorr, SwitchRand, SwitchWeight, and UniMin).}
 \label{fig:mf2}
\end{figure}

We next tested models of Bimanual proprioception for their ability to
predict observed bimanual proprioceptive bias and acuity. These data
are summarized in Table \ref{table:hypotheses} and displayed
graphically in Figures~\ref{fig:mf2}A~and~\ref{fig:mf2}B (``model
predictions''). Bimanual bias predicted from all three signal
combination models ($H_{BiMLE}$, $H_{BiMLEcorr}$, $H_{BiEqual}$) was
consistent with the observed data (p$>$0.05 in all cases). Predicted
acuity from $H_{BiMLE}$ and $H_{BiEqual}$ was observed to be reliably
better than empirically observed Bimanual proprioceptive acuity (p $<$
0.0001 in both cases). The hypothesis $H_{BiMLEcorr}$ adjusting MLE
for correlation between left and right limbs (using an estimate of
groupwise correlation $\rho$ between Left and Right $\rho = 0.33$)
predicted poorer proprioceptive acuity compared to $H_{MLE}$ (9.00
$pm$ 0.37~mm), but still predicted better acuity than the observed
empirical data (p=0.031). We also investigated the hypothesis that
correlation between unimanual signals may be subject-specific, and in
this case fit $\rho$ on a per-subject basis (rather than using the
groupwise $\rho = 0.33$ above). In this case predicted proprioceptive
acuity in this case was still better than empirically observed
Bimanual acuity (p = 0.034).

We next investigated the hypotheses which predict that subjects switch
between unimanual proprioceptive signals. Predicted proprioceptive
biases from the $H_{SwitchRand}$ and $H_{SwitchWeight}$ models were
not reliably different from empirically estimated Bimanual data (p$>$
0.05 in both cases). Proprioceptive acuity predicted by the
$H_{SwitchRand}$ model was reliably different (predicting better) from
the empirical data (p=0.031), while acuity predicted from the
$H_{SwitchWeight}$ model was also reliably better than that observed
empirically (p=0.0045).

One might also hypothesize that a single limb alone is used for
bimanual responses. Based on the above data showing reliable bias
differences between Left, Right and Bimanual responses it is clear
that subjects do not solely use the left or right limb exclusively for
Bimanual proprioception. It might instead be hypothesized that
subjects use the limb having the best proprioceptive acuity (the limb
with the smallest $\sigma$), a hypothesis we label $H_{UniMin}$. In
fact, bias and acuity predicted by $H_{UniMin}$ were consistent with
the observed Bimanual bias (p$=$0.28) and acuity (p$=$0.97). Of our 37
subjects, 22 (59$\%$) had best acuity with the Left hand.

\begin{figure}[p]
  \begin{center}
    \includegraphics[height=15cm]{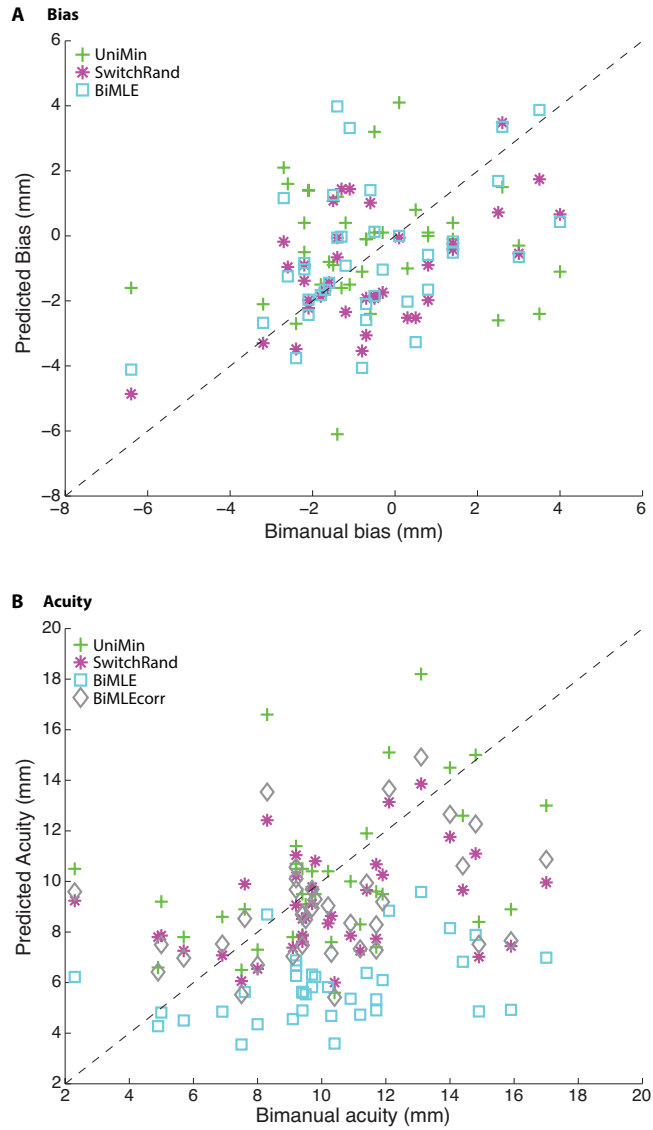}
  \end{center}
  \caption{A: Predicted bimanual bias as a function of empirically
    measured Bimanual bias, plotted for each subject. B: Predicted
    bimanual acuity as a function of empirically measured Bimanual
    acuity, plotted for each subject.}
  \label{fig:mf3}
\end{figure}

Figures~\ref{fig:mf3}A~and~\ref{fig:mf3}B plot individual subject data
showing predicted bias (Fig. \ref{fig:mf3}A) and acuity
(Fig. \ref{fig:mf3}B) for the models tested as a function of
empirically estimated values. Notably $H_{BiMLE}$ clearly
overestimates Bimanual acuity (i.e. underestimates $\sigma$).

\section*{Discussion}

This study examined sensory integration for proprioception of the two
limbs. The empirical data are consistent with the hypothesis that the
nervous system is aware of and uses the limb with the best
proprioceptive acuity for bimanual judgments. Our data are not
consistent with the prevailing model that predicts that the nervous
system optimally combines sensory signals from the two limbs. In fact
the maximum-likelihood model was worst at predicting bimanual acuity,
and adjustments made for signal correlation only slightly improved the
model's predictions, which were still reliably different (better) than
empirically observed bimanual acuity.

Why did participants not optimally combine proprioceptive signals from
the left and right limbs? It may be that the particular task tested
here is one for which the sensorimotor system does not have extensive
experience. For example it has been shown that in the absence of
practice the human sensorimotor system is not able to optimally
combine multiple sources of visual information for behaviours such as
navigation \cite{Souman:09}.

In our study the model with the most support was one that assumes
subjects know in advance which limb has the best proprioceptive
acuity. There is some support for the idea that the human sensorimotor
system maintains a representation of motor variability for left and
right limbs and uses this information both for online correction and
for planning of subsequent movement. The motor system makes
trial-by-trial adjustments to left and right limb trajectories during
bimanual reaching movements and that such adjustments are
preferentially made to movements of the non-dominant hand
\cite{White:10}. The non-dominant hand is in general less accurate
during reaching movements and it has thus been proposed that this
acuity difference causes the motor system to selectively adjust the
control signals for the less-accurate limb. Since our subjects were
right handed and Left and Right limb biases were both different from
Bimanual bias, these data are not consistent with either a dominant or
non-dominant hand hypothesis for bimanual proprioception. It is
certainly true that our task does not involve active movement, and in
fact several studies have shown that during static proprioception
(followed by an active matching movement) the majority of subjects are
more accurate at static limb proprioception with the non-dominant hand
\cite{Sainburg:02,Goble:06,Goble:08a,Goble:08b}.

A criticism of the current study might be related to a potential
cognitive component inherent in the psychophysical testing
procedure. It may be argued that proprioception could be similar to
the visual system with respect to its two-streams hypothesis for
perception and action
\cite{Volpe:79,Paillard:83,Rossetti:95,Dijkerman:07}. Visual
information for active movement has been shown to be distinct from
visual information for perception \cite{Schneider:69,Goodale:91}. If
such a dissociation exists in the somatosensory system, it may be that
signals from the left and right arms are integrated or combined
differently for a task that is less ``perceptual'' (such as the task
we used) and more ``dorsal'' in nature. Recently however the
double-dissociation hypothesis for the sense of somatosensation -
haptic touch specifically - was tested directly in a vibrotactile
experiment \cite{Harris:04}. Experimenters fit responses of normal
subjects to different signal detection models to determine whether
psychophysical responses could be explained by independent parallel
processes, or serial processes. Only the serial model successfully
described subject responses, leading the authors to conclude that
somatosensation for action and perception are not mutually independent
processes but rather localization is subsequent to detection. This
study illustrates that the two-streams hypothesis may not apply to
somatosensory function.

Bayesian predictions of sensory integration within the nervous system
face the scientific challenge that such predictions are consistent
with performance in any task for which optimal performance is
observed. That is, Bayesian models are a sufficient way of arriving at
optimal performance, but it is unclear if they are necessary. Several
recent studies have shown that the sensorimotor system's behaviour is
not always consistent with Bayesian predictions. Two recent studies
suggest that the motor system does not always have an accurate
estimate of its own motor variability, \cite{Zhang:13,Mamassian:08} a
prerequisite for Bayesian integration of information and the cause of
larger-than-optimal pointing errors.

Future studies may wish to attempt to incorporate some measure of
signal correlation in their experimental design. This would alleviate
the necessity of analytically fitting correlation coefficients as in
the current paper. On the other hand this would involve some
experimental challenges. It is not clear if subjects are capable of
accurately reporting the position of both hands simultaneously on each
psychophysical test, or if such simultaneous responses would
accurately reflect the trial-to-trial correlation of unimanual
proprioception.  \clearpage

\newpage
\section*{Acknowledgements}
This research was funded by Canadian Institutes of Health Research
(CIHR) Grant (PLG), The Netherlands Organization for Scientific
Research (DAK), a National Sciences and Engineering Council of Canada
graduate scholarship (ETW), and a Canada Graduate scholarship from
CIHR (JDW).

\newpage
\bibliographystyle{unsrt} 
\bibliography{refs}

\begin{thebibliography}{10}

\bibitem{Kording:04a}
Konrad~P K{\"o}rding and Daniel~M Wolpert.
\newblock Bayesian integration in sensorimotor learning.
\newblock {\em Nature}, 427(6971):244--7, Jan 2004.

\bibitem{Ernst:02}
Marc~O Ernst and Martin~S Banks.
\newblock Humans integrate visual and haptic information in a statistically
  optimal fashion.
\newblock {\em Nature}, 415(6870):429--33, Jan 2002.

\bibitem{Gweon:10}
Hyowon Gweon, Joshua~B Tenenbaum, and Laura~E Schulz.
\newblock Infants consider both the sample and the sampling process in
  inductive generalization.
\newblock {\em Proc Natl Acad Sci USA}, 107(20):9066--71, May 2010.

\bibitem{Bannard:09}
Colin Bannard, Elena Lieven, and Michael Tomasello.
\newblock Modeling children's early grammatical knowledge.
\newblock {\em Proc Natl Acad Sci USA}, 106(41):17284--9, Oct 2009.

\bibitem{Frank:12}
Michael~C Frank and Noah~D Goodman.
\newblock Predicting pragmatic reasoning in language games.
\newblock {\em Science}, 336(6084):998, May 2012.

\bibitem{Peterson:12}
Matthew~F Peterson and Miguel~P Eckstein.
\newblock Looking just below the eyes is optimal across face recognition tasks.
\newblock {\em Proc Natl Acad Sci USA}, 109(48):E3314--23, Nov 2012.

\bibitem{Cicchini:12}
Guido~Marco Cicchini, Roberto Arrighi, Luca Cecchetti, Marco Giusti, and
  David~C Burr.
\newblock Optimal encoding of interval timing in expert percussionists.
\newblock {\em J Neurosci}, 32(3):1056--60, Jan 2012.

\bibitem{Squeri:12}
Valentina Squeri, Alessandra Sciutti, Monica Gori, Lorenzo Masia, Giulio
  Sandini, and Juergen Konczak.
\newblock Two hands, one perception: how bimanual haptic information is
  combined by the brain.
\newblock {\em J Neurophysiol}, 107(2):544--50, Jan 2012.

\bibitem{Zhang:13}
Hang Zhang, Nathaniel~D Daw, and Laurence~T Maloney.
\newblock Testing whether humans have an accurate model of their own motor
  uncertainty in a speeded reaching task.
\newblock {\em PLoS Comput Biol}, 9(5):e1003080, May 2013.

\bibitem{Goble:08b}
D~Goble and S~Brown.
\newblock The biological and behavioral basis of upper limb asymmetries in
  sensorimotor performance.
\newblock {\em Neuroscience {\&} Biobehavioral Reviews}, 32(3):598--610, Jan
  2008.

\bibitem{Wilson:10}
Elizabeth~T Wilson, Jeremy Wong, and Paul~L Gribble.
\newblock Mapping proprioception across a 2d horizontal workspace.
\newblock {\em PLoS ONE}, 5(7):e11851, Jan 2010.

\bibitem{Ostry:10}
David~J Ostry, Mohammad Darainy, Andrew A~G Mattar, Jeremy Wong, and Paul~L
  Gribble.
\newblock Somatosensory plasticity and motor learning.
\newblock {\em J Neurosci}, 30(15):5384--93, Apr 2010.

\bibitem{Wong:11}
Jeremy~D Wong, Elizabeth~T Wilson, and Paul~L Gribble.
\newblock Spatially selective enhancement of proprioceptive acuity following
  motor learning.
\newblock {\em J Neurophysiol}, 105(5):2512--21, May 2011.

\bibitem{Oruc:03}
Ipek Oru{\c c}, Laurence~T Maloney, and Michael~S Landy.
\newblock Weighted linear cue combination with possibly correlated error.
\newblock {\em Vision Res}, 43(23):2451--68, Oct 2003.

\bibitem{Souman:09}
Jan~L Souman, Ilja Frissen, Manish~N Sreenivasa, and Marc~O Ernst.
\newblock Walking straight into circles.
\newblock {\em Curr Biol}, 19(18):1538--42, Sep 2009.

\bibitem{White:10}
Olivier White and J{\"o}rn Diedrichsen.
\newblock Responsibility assignment in redundant systems.
\newblock {\em Curr Biol}, 20(14):1290--5, Jul 2010.

\bibitem{Sainburg:02}
Robert~L Sainburg.
\newblock Evidence for a dynamic-dominance hypothesis of handedness.
\newblock {\em Exp Brain Res}, 142(2):241--58, Jan 2002.

\bibitem{Goble:06}
Daniel~J Goble, Colleen~A Lewis, and Susan~H Brown.
\newblock Upper limb asymmetries in the utilization of proprioceptive feedback.
\newblock {\em Exp Brain Res}, 168(1-2):307--311, Jan 2006.

\bibitem{Goble:08a}
D.~J Goble and S.~H Brown.
\newblock Upper limb asymmetries in the matching of proprioceptive versus
  visual targets.
\newblock {\em J Neurophysiol}, 99(6):3063--3074, Apr 2008.

\bibitem{Volpe:79}
B~T Volpe, J~E LeDoux, and M~S Gazzaniga.
\newblock Spatially oriented movements in the absence of proprioception.
\newblock {\em Neurology}, 29(9 Pt 1):1309--13, Sep 1979.

\bibitem{Paillard:83}
J~Paillard, F~Michel, and G~Stelmach.
\newblock Localization without content. a tactile analogue of 'blind sight'.
\newblock {\em Arch Neurol}, 40(9):548--51, Sep 1983.

\bibitem{Rossetti:95}
Y~Rossetti, G~Rode, and D~Boisson.
\newblock Implicit processing of somaesthetic information: a dissociation
  between where and how?
\newblock {\em Neuroreport}, 6(3):506--10, Feb 1995.

\bibitem{Dijkerman:07}
H.~Chris Dijkerman and Edward H. F~De Haan.
\newblock Somatosensory processes subserving perception and action.
\newblock {\em Behav. Brain Sci.}, 30(02):189, Apr 2007.

\bibitem{Schneider:69}
G~E Schneider.
\newblock Two visual systems.
\newblock {\em Science}, 163(3870):895--902, Feb 1969.

\bibitem{Goodale:91}
M~A Goodale, A~D Milner, L~S Jakobson, and D~P Carey.
\newblock A neurological dissociation between perceiving objects and grasping
  them.
\newblock {\em Nature}, 349(6305):154--6, Jan 1991.

\bibitem{Harris:04}
Justin~A Harris, Thida Thein, and Colin W~G Clifford.
\newblock Dissociating detection from localization of tactile stimuli.
\newblock {\em J Neurosci}, 24(14):3683--93, Apr 2004.

\bibitem{Mamassian:08}
Pascal Mamassian.
\newblock Overconfidence in an objective anticipatory motor task.
\newblock {\em Psychol Sci}, 19(6):601--6, Jun 2008.

\end{thebibliography}

\begin{table}
  \centering
  \begin{tabular}{|c|c|c|c|c|c|}
    \hline
    \multicolumn{2}{|c}{} & \multicolumn{2}{|c|}{Bias (mm)} & \multicolumn{2}{|c|}{Acuity (mm)} \\ 
    \cline{3-6}
    \multicolumn{2}{|c|}{} & est & p & est & p\\ 
    \cline{1-6}
    \multicolumn{2}{|c|}{Empirical} & 0.60 $\pm$ 0.34 &  & 10.15 $\pm$ 0.52 & \\ 
    \hline
    \multirow{3}{*}{Bimanual models} & BiEQ & 1.01 $\pm$ 0.29 & $>$0.05 & 7.89 $\pm$ 0.32 & $<$0.05\\ 
    & BiMLE & 0.65 $\pm$ 0.34 & $>$0.05 & 5.85 $\pm$ 0.23 & $<$0.001\\ 
    & BiMLECorr & 0.65 $\pm$ 0.34 & $>$0.05 & 9.00 $\pm$ 0.37 & $<$0.001\\ 
    \hline
    \multirow{3}{*}{Unimanual} & SwitchRand & 1.05 $\pm$ 0.29 & $>$0.05 & 9.04 $\pm$ 0.32 & $<$0.05\\ 
    & SwitchWeight & 0.70 $\pm$ 0.34 & $>$0.05 & 8.62 $\pm$ 0.31 & $<$0.01\\ 
    & UniMin & 0.36 $\pm$ 0.30 & $>$0.05 & 10.14 $\pm$ 0.47 & $>$0.05\\ 
    \hline
  \end{tabular}
  \caption{Summary of mean $\pm$ SE of empirically estimated and predicted
    bimanual bias and acuity. Statistical tests (paired t-tests) were performed
    to test for reliable differences between empirical bimanual data and
    predictions.}
  \label{table:hypotheses}
\end{table}

\newpage

\newpage

\newpage

\end{document}